\PassOptionsToClass{
reprint,
superscriptaddress,
amsmath, amssymb,
aps,
prl,
floatfix
}{revtex4-2}
\documentclass{revtex4-2}

\usepackage[table]{xcolor}
\usepackage{graphicx}%
\usepackage{dcolumn}%
\usepackage{bm}%
\usepackage{enumitem} %
\usepackage[colorlinks=true, allcolors=blue]{hyperref}%
\usepackage[normalem]{ulem} %

\usepackage{bibunits}
\defaultbibliography{bibliography}
\defaultbibliographystyle{apsrev4-2}

\setcitestyle{super}
\usepackage[T1]{fontenc}
\usepackage{newtxtext,newtxmath}

\graphicspath{{figures/}}
\usepackage[separate-uncertainty=false]{siunitx}
\usepackage{braket}

\usepackage[capitalise]{cleveref}

\begin{document}
\begin{bibunit}

\title{Emerging supersolidity from a polariton condensate in a photonic crystal waveguide}
\date{\today}

\author{Dimitrios Trypogeorgos}
\email[]{dimitrios.trypogeorgos@nanotec.cnr.it}
\affiliation{CNR Nanotec, Institute of Nanotechnology, via Monteroni, 73100, Lecce, Italy}

\author{Antonio Gianfrate}
\affiliation{CNR Nanotec, Institute of Nanotechnology, via Monteroni, 73100, Lecce, Italy}

\author{Manuele Landini}
\affiliation{Institut f\"ur Experimentalphysik und Zentrum f\"ur Quantenphysik, Universit\"at Innsbruck, 6020 Innsbruck, Austria}

\author{Davide Nigro}
\author{Dario Gerace}
\affiliation{Dipartimento di Fisica, Unversit\`a degli Studi di Pavia, via Bassi 6, 27100 Pavia, Italy}

\author{Iacopo Carusotto}
\affiliation{Pitaevskii BEC Center, CNR-INO and Dipartimento di Fisica, Universit\`a di Trento, I-38123 Trento, Italy}

\author{Fabrizio Riminucci}
\affiliation{Molecular Foundry, Lawrence Berkeley National Laboratory, One Cyclotron Road, Berkeley, California, 94720, USA}

\author{Kirk W. Baldwin}
\author{Loren N. Pfeiffer}
\affiliation{PRISM, Princeton Institute for the Science and Technology of Materials, Princeton University, Princeton, New Jersey 08540, USA}

\author{Giovanni I. Martone}
\affiliation{INFN, Sezione di Lecce, 73100 Lecce, Italy}
\affiliation{CNR Nanotec, Institute of Nanotechnology, via Monteroni, 73100, Lecce, Italy}

\author{Milena De Giorgi}
\author{Dario Ballarini}
\author{Daniele Sanvitto}
\affiliation{CNR Nanotec, Institute of Nanotechnology, via Monteroni, 73100, Lecce, Italy}

\begin{abstract}
A supersolid is a counter-intuitive phase of matter where its constituent particles are arranged into a crystalline structure, yet they are free to flow without friction.
This requires the particles to share a global macroscopic phase while being able to reduce their total energy by spontaneous, spatial self-organisation.
This exotic state of matter has been achieved in different systems using Bose-Einstein condensates coupled to cavities, possessing spin-orbit coupling, or dipolar interactions.
Here we provide experimental evidence of a new implementation of the supersolid phase in a novel non-equilibrium context based on exciton-polaritons condensed in a topologically non-trivial, bound-in-the-continuum state with exceptionally low losses.
We measure the density modulation of the polaritonic state indicating the breaking of translational symmetry with a remarkable precision of a few parts in a thousand.
Direct access to the phase of the wavefunction allows us to additionally measure the local coherence of the superfluid component.
We demonstrate the potential of our synthetic photonic material to host phonon dynamics and a multimode excitation spectrum.
\end{abstract}

\maketitle

The existence of a supersolid phase of matter, combining a crystalline structure with superfluid properties, was speculated more than 50 years ago~\cite{PhysRev.106.161,1969JETP...29.1107A,PhysRevLett.25.1543,1971JETP...32.1191K,THOULESS1969403,GROSS195857} however only recently there have been convincing experimental evidence, mainly using ultracold atomic Bose-Einstein condensates (BECs) coupled to electromagnetic fields.
There, various guises of the supersolid were created using atoms coupled to high-finesse cavities~\cite{leonard.nature.2017,gopalakrishnan.nphys.2009,guo.nature.2021}, with large magnetic dipole moments~\cite{tanzi.nature.2019,tanzi.science.2021,PhysRevLett.122.130405,PhysRevX.9.021012,bottcher.prx.2019,klaus.nphys.2022,guo.nature.2019,bottcher.prx.2019}, and spin-orbit-coupled, two-component systems showing stripe phases~\cite{li.nature.2017,putra.PhysRevLett.124.053605}.

At the mean field level, supersolidity can be interpreted as two-mode condensation; after condensation in the first, a second mode becomes energetically available by tuning of interactions or an external electromagnetic field and can then be dynamically populated.
In a single-mode BEC, the U(1) symmetry related to its phase is broken at the condensation phase transition, giving rise to superfluidity in the system.
The mean field wavefunction has $k=0$ momentum, giving a constant density in space.
Supersolidity arises, in most cases, as an additional, macroscopic occupation of one (or more) finite-momentum modes of the system leading to breaking of translational symmetry.
The occupation is preceded by the presence of a roton-like minimum in the excitation spectrum that can be dynamically populated when its energy is tuned to be equal to that of the stationary ground state~\cite{chomaz.nphys.2018,PhysRevA.90.063624,PhysRevLett.114.105301}.
The new mean field wavefunction acquires therefore density modulations with a given length scale and a macroscopic phase relation, giving rise to an extra U(1) symmetry breaking.
A change in the relative phase of the two condensates corresponds to a rigid translation of the density pattern associated to a Goldstone mode of the system~\cite{wouters.goldstone.PhysRevA.76.043807}.
A gapped Higgs mode is also present, corresponding to a change in the amplitude of the density modulation~\cite{leonard.science.aan2608}.

However, despite all the progress in understanding the physics of the supersolid phase, most current realisations are limited to ultracold atomic systems.
Here we demonstrate a new photonic platform that can give rise to a supersolid phase in a novel driven-dissipative context and is poised to open new directions for the study of these intriguing phases of matter.
We use a BEC of polaritons, i.e., strongly coupled light-matter excitations, formed in a sub-wavelength, patterned waveguide that gives rise to a condensate at a saddle point of the dispersion~\cite{ardizzonePolaritonBoseEinstein2022,kasprzak_boseeinstein_2006,carusotto_quantum_2013,amo_superfluidity_2009,sanvitto_all-optical_2011}.
Under some mode-folding conditions and exciton-photon detuning, the photonic crystal waveguide shows two extra modes symmetrically distributed in $k$ with respect to the BEC at $k=0$, with the same polarisation.
This allows for the formation of energy-degenerate, optical parametric oscillation (OPO) between the condensate and the adjacent modes, which can be described in terms of polariton-polariton interactions arising from a $\chi^{(3)}$ nonlinearity.
Parametric processes are commonplace in nonlinear photonics and can exhibit pattern formation close to threshold in multimode systems~\cite{pattern.formation.PhysRevLett.83.5278,ardizzone2013.scientific.reports}.
However, differently from our results, they are usually induced resonantly which imposes the coherence of the pump laser on the wavefunction~\cite{whittaker_polariton_2017,romanelli_four_2007,lewandowski_polarization_2016}.
Recently, the onset of a non-resonantly-driven, energy-degenerate parametric scattering was demonstrated in different systems~\cite{wu_nonlinear_2021,sawicki_polariton_2021,PhysRevB.87.155302,doi:10.1021/acs.nanolett.1c04800,PhysRevB.107.L060303,PhysRevLett.108.166401} showing how these scattering processes are ubiquitous; however, no proof of coherence nor pattern formation was reported.
Here, instead, we used a $\Gamma$-point, symmetry protected, bound-in-the-continuum (BiC) state formed at the lower branch of the anticrossing~\cite{ardizzonePolaritonBoseEinstein2022,gianfrateReconfigurableQuantumFluid2024}, which led to a few crucial advantages: (1) a vanishing linewidth due to the supression of losses on the photonic component, which reduces both the condensation threshold and the incoherently scattered background, and does not overwhelm the emission from other modes in the system, (2) a negative effective mass, which makes the $k=0$ component self-localised, and (3) absence of accessible lower-energy extrema in the dispersion where particles can accumulate; as such, the OPO is adequately described by a three-branch iso-energetic model.
 These striking properties of the BiC render the small density modulation of the BEC clearly visible.

\begin{figure*}[htb]
	\begin{center}
	\includegraphics[width=2\columnwidth]{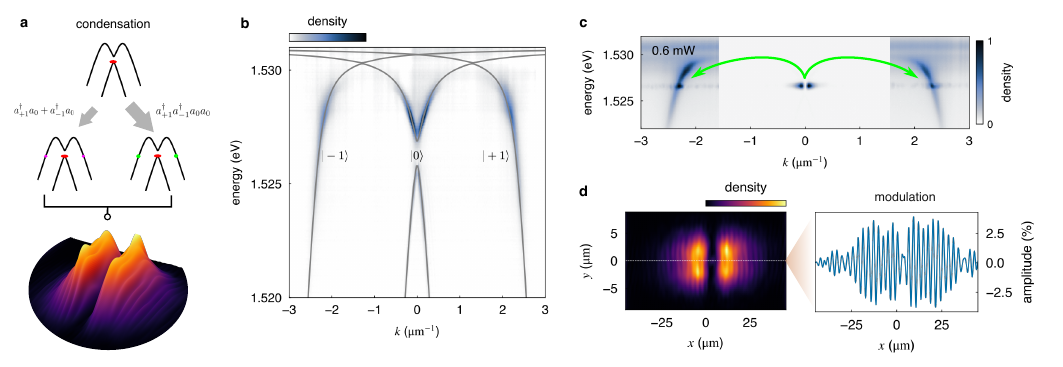}
	\caption[]{\textbf{System in reciprocal and coordinate space.} \textbf{a.} The mechanism that leads to the formation of the supersolid. Linear and nonlinear scattering processes combine and give rise to a density modulation in coordinate space. \textbf{b.} The dispersion comprises of four fundamental propagating modes. The measured dispersion closely matches the numerical dispersion obtained from diagonalising an effective $8\times8$ Hamiltonian (see SI). \textbf{c.} The BiC state $\ket{0}$ at $k=0$ parametrically scatters to the modes $\pm k_r$, denoted as $\ket{\pm 1}$. At powers above threshold, the BEC needs to be masked in order for the $\ket{\pm 1}$ population to be measured. The parametric scattering generates photon pairs that populate the $\ket{\pm 1}$ states as indicated by the arrows. \textbf{d.} The density of the supersolid wavefunction $\psi(x,y)$ shows a characteristic fast modulation due to the spontaneous breaking of translational symmetry. The amplitude of the modulation along the $y=0$ line is $\sim$2.6\%.
 }
	\label{fig:1}
	\end{center}
\end{figure*}

\section{Formation of the density modulation}

We prepared a polariton BEC, using off-resonant, pulsed excitation, at the lower branch of an anticrossing formed by folding the propagating modes of the one-dimensional waveguide to $k=0$ using a sub-wavelength, etched grating.
The corresponding wavevector sets the size of the irreducible Brillouin zone, which is much larger than all other characteristic length scales of the system, and lies outside our numerical aperture (see Methods).
Aside from the BiC, which corresponds to the anti-crossing of the left- and right-propagating TE$_0$ modes, the waveguide supports also TE$_1$ modes with the mode spacing determined by the planar waveguide thickness and index contrast, which are responsible for the formation of the modulation.
The off-resonant pumping predominantly populates the BiC due to its negative effective mass~\cite{riminucciPolaritonCondensationGapConfined2023,nigro.PhysRevB.108.085305}.
It is important to note that the TE$_1$ modes are populated by some incoherent population relaxing from the exciton reservoir, however they can also become macroscopically populated via linear and in particular nonlinear coupling with the TE$_0$ (\cref{fig:1}a).

The mode-spacing sets the natural unit of length for our system to be $1/k_r=\qty{2.3}{\micro m}$, so we can write the four basis states as $\ket{\pm, mk_r}$ for $m=0,1$, where the sign labels the propagation direction of the relevant photonic dispersion branch. %
The TE$_0$ modes are coupled with a Rabi frequency of \qty{1.5}{meV}, %
and the BiC corresponds to the lowest energy adiabatic state $\ket{0}=(\ket{+,0}+\ket{-,0})/\sqrt{2}$ %
that has zero group velocity, with an energy $E = \qty{1.526}{eV}$. We will from now on limit the description to the three states $\ket{0},\ket{\pm1}\equiv\ket{\pm,k_r}$ to simplify notation.
The single-particle dispersion shown in \cref{fig:1}b is measured below threshold since the formation of the BEC dominates the luminescence above threshold and renders the rest of the dispersion scarcely visible.
The BEC is visible in the dispersion at the $\Gamma$-point, slightly blueshifted from the bare dispersion due to intrinsic polariton-polariton interactions, \cref{fig:1}c~\cite{ardizzonePolaritonBoseEinstein2022}.
The excitonic Hopfield coefficient at the BiC energy is estimated to be 0.54 for $\ket{0}$ and 0.4 for $\ket{\pm 1}$, due to their different mode structure that leads to them coupling to the exciton with different Rabi frequencies.

We will assume, in what follows, no linear coupling between TE$_0$ and TE$_1$ modes. i.e. the off-diagonal element in the hamiltonian $U_{01}\to 0$ (see SI).
When the population of $\ket{0}$ becomes large enough, parametric scattering driven by a four-polariton interaction term of the form $a_{+1}^\dagger a_{-1}^\dagger a_0 a_0+h.c.$, where $a_m$ are bosonic operators, results in emission of polariton pairs that populate the $\ket{\pm 1}$ states at finite momentum according to energy and momentum conservation criteria. The $\ket{0}$ and $\ket{\pm 1}$ states are commonly referred to as pump, signal, and idler modes in optical parametric processes.
Since the interaction term is invariant to the transformation $a_m \to a_m e^{mi\phi}$, with $m=0,\,\pm1$~\cite{wouters.PhysRevA.76.043807}, it has U(1) symmetry which is spontaneously broken at OPO threshold.
At the same time the OPO breaks translational symmetry, due to it being energy-degenerate, which leads to the appearance of a density modulation on top of the condensate~\cite{carusotto.PhysRevB.72.125335}.

Thanks to the small radiative coupling of the BiC with the far field, the luminescence of the BEC is strongly suppressed at $k=0$, showing the characteristic two-lobe structure with a $\pi$ phase jump between them (\cref{fig:1}d).
This is essentially due to the co- and counter-propagating TE$_0$ modes being out of phase at $k=0$ as they Bragg scatter from the grating; the propagation into the far field leads to the characteristic nodal line at the centre of the BEC.
The density modulation, indicating the translational symmetry breaking of the supersolid, is visible atop both lobes of the BEC, which are extended in reciprocal space and outside the laser spot in coordinate space.
Its amplitude is shown in \cref{fig:1}e to be $A\sim$2.6\%.
The modulation pattern extends over $2k_r L \simeq 30$ lattices sites, where $L=\qty{30}{\mu m}$ is the effective $1/e^2$  waist of the BEC.
The presence of the modulation on top of the BEC is a clear sign of crystallisation: the observation of a periodic pattern, similar to the formation of crystals, despite the system otherwise still being in the superfluid regime.

The density in \cref{fig:1}d is the integrated emission over approximately \num{8e6} pulses exciting the sample at a rate of \qty{80}{MHz}, much slower than the lifetime of the BEC (of the order of a few hundred picoseconds); as such, each measurement is an average of multiple independent BEC implementations.
The dynamical behaviour of the optical parametric process near- and above-threshold can be simply described by a multibranch Hamiltonian involving three modes~\cite{PhysRevB.87.155302,langbein.PhysRevB.70.205301,ciuti.PhysRevB.63.041303}.
The single-mode solution becomes unstable with increasing power and a three-mode solution becomes the ground state of the system~\cite{dunnett.PhysRevB.98.165307,PhysRevB.87.155302}.

\begin{figure}[tb]
	\begin{center}
	 \includegraphics[width=\columnwidth]{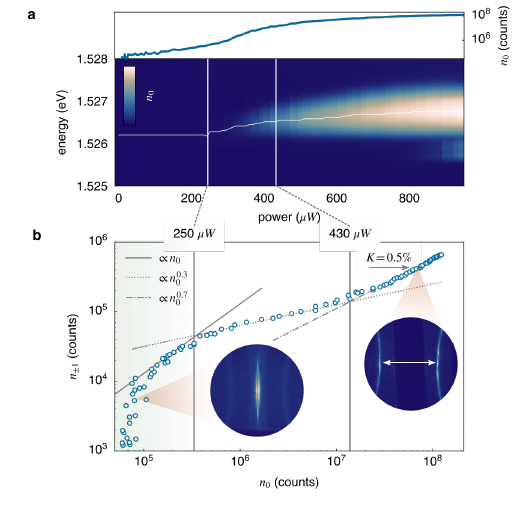}
	\caption[]{\textbf{Parametric scattering.} \textbf{a.} Growth of the BEC at $k=0$. Above the condensation threshold power \qty{250}{\micro W} both the blueshift and linewidth start increasing to about \qty{1}{meV}, typical of pulsed excitation. The first-excited mode appears for powers above \qty{800}{\micro W} at the energy of the lower band. \textbf{b.} Emission from $k=0$ and $k=k_r$ that are proportional to the poalriton BEC density. Coherent and incoherent scattering processes can be separated since they have different growth behaviour with the BEC density. The $n_{\pm 1}$ density experiences sublinear growth above threshold that then shows a second threshold in efficiency at about $\times 1.72$ above threshold power. The insets show how the distribution of the density on the TE$_1$ modes changes from uniform at low powers to more localised along $k_x$ above the second threshold.
 }
	\label{fig:2}
	\end{center}
\end{figure}

Figure~\ref{fig:2}a
shows that condensation takes place at around \qty{0.3}{mW} of pump power.
The luminescence and linewidth of the BEC dramatically increase above threshold due to the pulsed excitation.
Since our system is trapped, it supports multiple confined modes; the first excited mode enters in the gap around \qty{0.8}{mW}, and can be clearly seen in \cref{fig:2}a at lower energies than the main, blueshifted BEC~\cite{riminucciPolaritonCondensationGapConfined2023}.
Figure~\ref{fig:2}b shows the growth of the densities $n_{\pm 1}$ versus $n_0$, showing approximate linear growth below BEC threshold.
This suggests that at low excitation power, the emission is dominated by incoherent photoluminescence coming from the exciton reservoir.
The $\ket{0}$ mode is populated much more efficiently and $n_{\pm 1}$ starts to experience sublinear growth above threshold.
At roughly \qty{430}{\mu W},
the efficiency of the scattering process increases indicating that nonlinear scattering is taking place.
In this region the $\ket{\pm 1 }$ states, that are uniformly populated in reciprocal space for low pump powers, become structured showing conjugate scattered pairs (see insets of \cref{fig:2}b) and the density modulation becomes even more evident in coordinate space.

Our ability to observe the density modulation comes from the small linear coupling present in the system.
Since the $\ket{\pm1}$ states are not empty when the OPO process establishes, but are populated by the exciton reservoir or via Rayleigh scattering, even a small amount of linear coupling contributes towards fixing the phase of the density modulation and as such renders it visible in time-integrated emission.
The linear coupling is equivalent to including two-operator exchange terms of the form $a_{+1}^\dagger a_0 + a_{-1}^\dagger a_0 + h.c.$ in the hamiltonian, which make it not invariant under the phase transformation discussed above; formally the OPO tends to break the U(1) symmetry in a non fully spontaneous manner because of this.
Well above the OPO threshold the nonlinear scattering dominates and a degree of phase freedom is recovered.
To show this we consider the ratio $K = (n_1 + n_{-1}) / 2n_0 $, which is effectively the Bragg signal corresponding to a given modulation amplitude, and scales as $A^2/4$,
where $1/4$ is the structure factor for a sinusoidal modulation.
For our observed density modulation, this gives an expected Bragg signal of 0.02\%, more than an order of magnitude smaller than the measured 0.5\%.
Moreover, we note that (1) the phase of the modulation varies in the transverse direction by about a $1.1\pi$ rotation, as seen in \cref{fig:1}d, and (2) the average phase in continuous-wave excitation fluctuates predominantly at the density maxima (see discussion in SI).
These observations indicate underlying phase fluctuations related to the OPO that are averaged out during an integrated measurement and effectively reduce the visibility of the density modulation.
Analogous effects exist in atomic supersolids, where the phase of the modulation is fixed by the presence of an external trapping potential, yet small phase variations could still be observed, e.g. in dipolar supersolids~\cite{PhysRevLett.122.130405}.
We note the linear coupling term alone does not fix the phase of the modulation as discussed for spin-orbit-coupled systems~\cite{li.yun.PhysRevLett.108.225301,geier.PhysRevLett.130.156001}.
Such effects are inherent in finite-size systems, however they have not prevented the observation of well-behaved collective excitation spectra~\cite{tanzi.nature.2019,tanzi.science.2021}.
Importantly, our system can recover full phase freedom in a much simpler manner, i.e. by engineering the dispersion so that $U_{01} \to 0$.

\section{Local and global coherence}

\begin{figure}[htb]
	\begin{center}
	 \includegraphics[width=\columnwidth]{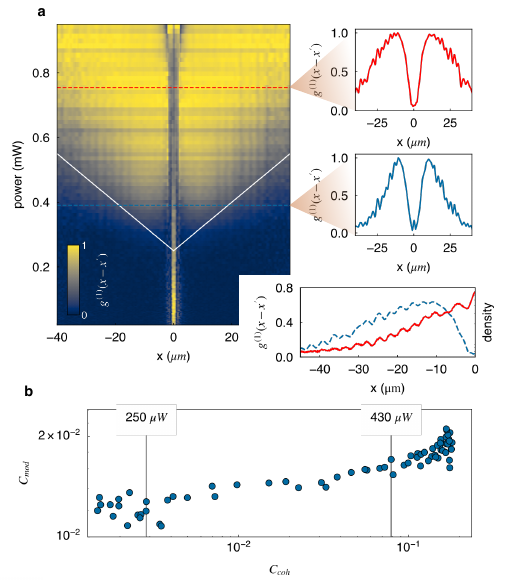}
	\caption[]{\textbf{Spatial coherence through threshold.} \textbf{a.} The first-order correlator $g^{(1)}(x-x^\prime)$ is a direct measurement of coherence of the superfluid part. Since it is normalised to unity at each power, the increase of the width of the distribution is proportional to the increase of the coherence length (see right-hand panels) marked by the white diagonal lines with a slope \qty{133}{\mu m/mW}. The modulations on the $g^{(1)}(x-x^\prime)$ are correlated with the density modulations showing a local drop of coherence at the individual crystal lattice sites (lower right panel at \qty{754}{\mu W}). \textbf{b.} Growth of $C_{mod}$ versus $C_{coh}$ over threshold. The trend is similar to \cref{fig:2}b showing the existence of two thresholds.}
	\label{fig:3}
	\end{center}
\end{figure}

Having shown the behaviour of the populations undergoing parametric scattering, it is interesting to see how the diagonal and off-diagonal phase coherence is established.
We use interferometric measurements to fully capture the microscopic, spatial coherence of the system, which serves as a direct proof that the fragile local and long-range order of the solid and superfluid parts is preserved.
The strength of using an all-optical system in this regard is in the ease of access to the first-order spatial correlator, $g^{(1)}(x-x^\prime)$, due to our ability to measure the full wavefunction of the BEC $\psi(x,y) = \sqrt{n(x,y)} \exp(-i \phi(x, y))$ using off-axis holography (see Methods).
Direct, local measurements of the first-order coherence show the wavefunction to be fully coherent above threshold with a modulated coherence amplitude locally correlated with the emerging supersolid crystalline structure.
The superfluid character of our bulk BEC on the BiC state has already been characterised through the observation of a linear Bogoliubov excitation spectrum~\cite{nina_collective_2023}, and as such, here we take the observed increase in the $g^{(1)}(x-x^\prime)$ coherence length (\cref{fig:3}a) to be indicative of an increased superfluid fraction.

Figure~\ref{fig:3}a shows the increase with power of $g^{(1)}(x-x^\prime)$ normalised to its maximum.
Below threshold only the the zero-time-space auto-correlator $g^{(1)}(0)$ has appreciable amplitude.
At \qty{\sim 250}{\mu W}, the onset of polariton condensation, the coherence length starts increasing at a rate of \qty{133}{\mu m/mW}.
The fast spatial oscillations that appear in $g^{(1)}(x-x^\prime)$ have an amplitude of $\sim$1.5\% and are in phase with the density modulations, showing how coherence locally drops in regions of lower density of the crystalline solid~\cite{PhysRevLett.25.1543} (see \cref{fig:3}a).

The emergence of the global, diagonal and off-diagonal long-range order can be quantified by measuring $\mathcal{F}_x\{\vert\psi(x, y)e^{-ik_1x} + \psi(-x, -y) e^{-ik_2x}\vert^2\}$ at $y=0$, where $\mathcal{F}_x\{\cdot\}$ is the spatial Fourier transform along $x$ and the quantity in the brackets is the output of the interferometer (see SI).
The magnitude of the $k=k_r$ component is a measure of the amplitude of the density modulation arising from the phase coherence between the three BEC components.
Similarly, the component corresponding to the wavevector of the interferogram fringes, $k=k_2 - k_1$, captures the phase coherence of the $k=0$ component.
We denote these quantities as $C_{mod}$ and $C_{coh}$ respectively.
The relative increase of $C_{mod}$ against $C_{coh}$, above threshold pump power, (see \cref{fig:3}b) is similar to the behaviour of $n_{\pm 1}$ against $n_0$ shown in \cref{fig:2}b.
Below the condensation threshold, $C_{mod}$ is largely dominated by noise since the crystalline pattern has not formed yet, then it slowly rises up to the OPO threshold where it begins to increase sharply.
In absence of linear coupling this would indicate a two-step breaking of the superfluid U(1) and translational symmetries; in our case the density modulation increasing is a clear signature that the OPO process is established.

\section{Non-rigidity of the modulation}

An important aspect of the study of quantum many-body systems is understanding their collective excitations; they dictate the dynamics of the system and its response to external perturbations.
In order to be able to observe and manipulate the collective modes, the quantum system needs to be allowed to spontaneously break the relevant symmetry during a phase transition, without external or intrinsic factors that influence this choice.
For example, in single-mode, high-finesse cavity realisations of the supersolid phase the wavelength of the modulation is fixed by the wavelength of the relevant cavity mode, implying that the phononic branch corresponding to the Goldstone mode cannot be observed~\cite{leonard.science.aan2608}.
In such systems, excitations at finite $k$ are suppressed due to the infinite range of the coupling.
This means that the global translation mode at $k=0$ exists, but the local compression/dilation motion of the fringes occurring at finite $k$ is not observable; such phonon
dynamics can be captured with multimode cavities or in stripe phases of spin-orbit-coupled systems, as well as dipolar systems due to the translational invariance of the interactions in those systems~\cite{guo.nature.2021,li.nature.2017,geier.PhysRevLett.130.156001}.
In our case, the symmetry breaking results from short range polariton-polariton interactions, we therefore expect similar behaviour to the latter systems.

\begin{figure}[htb]
	\begin{center}
	 \includegraphics[width=\columnwidth]{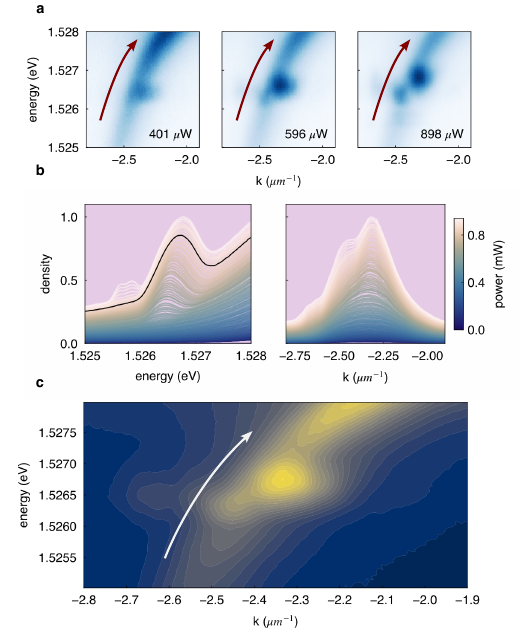}
	\caption[]{\textbf{Non-rigid scattering.} \textbf{a.} The energy and momentum of $\ket{-1}$ changes with excitation power. For high powers the first-excited mode is visible (right panel). The position of $\ket{-1}$ in the waveguide dispersion defines the interference part of the supersolid. Since the modes are curved due to them being strongly coupled to the exciton, the position moves both in energy \textbf{b.} (left) and in wavenumber (right) giving rise to a non-rigid lattice structure. The shift of the central peaks follows the dispersion curve shown in the top panels. The secondary peak appearing at lower energy is the first excited state that enters at $k=0$ for pump powers $>\qty{0.8}{mW}$ (marked by the black solid line) and scatters to the TE$_1$ modes.
 \textbf{c.} Energy dispersion averaged over all excitation powers shwoing the movement of $\ket{-1}$ in reciprocal space.
 }
	\label{fig:4}
	\end{center}
\end{figure}

The parametric scattering process is constrained to occur on the available $\ket{\pm 1}$ modes where the density of states is different from zero.
This fixes the wavenumber of $\ket{\pm 1}$ to $k_r$.
However the value of $k_r$ depends on polariton-polariton interactions and their resulting blueshift.
As the interactions blueshift the energy of the BEC, and since OPO is a iso-energetic process, the energy of $\ket{\pm 1}$ has to shift as well, and this is done by shifting their wavevector inwards.
This shift is nonlinear due to the strong-coupling of the photonic modes to the exciton, which leads to an increased curvature at energies closer to the exciton resonance.
The energy of $\ket{\pm 1}$, associated with the breaking of translational symmetry, increases by 0.35\,meV as the pump power increases (\cref{fig:4}a); the linewidth of the TE$_1$ mode is about \qty{500}{\mu eV}.
For the same range of pump powers $k_r$ shifts by \qty{0.04}{\micro m^{-1}} (\cref{fig:4}b, c).
As such, the wavevector of the density modulation is not fixed, but depends on both the single-particle and manybody parameters of the system, reminiscent of a non-rigid, roton-like  behaviour.

\section{Discussion and outlook}

We demonstrated evidence of a supersolid state of matter in a polaritonic system which is a novel and flexible platform for the investigation of the physics of supersolid systems.
To place our work more generally in the context of supersolids, as observed mainly in ultracold atomic systems, we experimentally addressed its characteristics in coordinate and reciprocal space, its diagonal and off-diagonal long range order, and its interaction-dependent behaviour.
Our supersolid state is formed using an iso-energetic OPO process to spontaneously break the translation symmetry of a non-resonant polaritonic superfluid.
We stress that this is a novel mechanismg for the creation of a supersolid, particular of the driven-dissipatve context of non-equilibrium polariton systems, and not simply a photonic analogue of mechanisms demonstrated in atomic platforms.
The analysis of the parametrically scattered light indicates that the system condenses at a given threshold and OPO activates at higher powers.
Theoretically, we also expect the existence of two dictinct transition points.

By exploiting the richness of engineering opportunities in these photonic crystal systems it is possible to eliminate completely the linear coupling between the modes, for example using TE- and TM-polarised modes.
This will enable us to observe full phase freedom of the density modulation either in single-shot experiments or higher-order coherence functions~\cite{carusotto.PhysRevB.72.125335,folling.spatial.coherence.nature.2005}.
Near and above threshold the statistics of the emitted light changes to bunched leading to an increase in density-density correlations that are peaked at threshold, similar to a resonantly pumped OPO~\cite{ardizzone_bunching_2012}.
A spatially dependent $g^{(2)}(x -x')$ will show signatures of density modulation since it is only sensitive to how polaritons at relative distances are correlated.

Since the spacing of the crystalline modulation is not rigid but influenced by short range polariton interactions, the low-lying excitation spectrum of the supersolid should reveal the two phononic Goldstone branches resulting from the symmetries involved.
The observation of such spectrum is a natural next step in the investigation of the supersolid properties.
As a further development of this research, increasing the dimensionality of the system, we expect additional breaking of discrete rotational symmetries to occur, as well as the proliferation and dynamics of stochastic vortices~\cite{pomeau.PhysRevLett.72.2426}.
Moving beyond parametric scattering, similar effects have been suggested utilising coupling to indirect excitons to engineer long-range, dipolar interactions that can lead to crystallisation under compression~\cite{boning.PhysRevB.84.075130}, or using Bose-Fermi mixtures where excitons are coupled to an electron gas~\cite{PhysRevLett.108.060401}.

\section{Methods}

\textbf{Sample.}
The sample used in these experiments is an AlGaAs waweguide grown with molecular beam epitaxy.
It has twelve 20\,nm embedded GaAs quantum wells with a heavy hole excitonic transition at \qty{1531.1(0.1)}{\milli \electronvolt}.
The propagating modes are folded within the lightcone by inscribing a $50\,\times\,400\,\unit{\mu m}$ unidirectional grating within the structure with a lattice parameter $a=\qty{242}{nm}$ corresponding to a wavevector translation of $\pi/a=\qty{13}{\mu m ^{-1}}$.
The energy momentum dispersion therefore consists of two counterpropagating TE$_0$ modes forming an anticrossing at $k_x=0$, \qty{2.4}{meV} blue detuned with respect to the excitonic transition and a couple of symmetric TE$_1$ modes located at $\approx \qty{2.3}{\mu m ^{-1}}$ from the TE$_0$ modes.
The filling factor of such grating is 76\% in order to spatially overlap the two $\ket{0}$ and ensure the coupling described by the $U$ term in the Supplemental information responsible of the opening of the BiC gap.
Due to the confinement, the strong coupling regime with a Rabi frequency of 5.6 and 4.2 \,meV is achieved between the excitonic transition and the  propagating TE$_0$ and TE$_1$ modes respectively.
The fabrication details of the grating etching can be found in~\cite{riminucci_nanostructured_2022}.

\textbf{Laser excitation scheme.}
The sample is held in an attoDRY closed loop helium cryostat at a temperature \qty{\approx 4}{\kelvin} in order to avoid exciton ionisation.
The waveguide is non-resonantly excited in reflection configuration using an \qty{80}{MHz} repetition rate, \qty{100}{\fs} tunable laser.
The polarisation is chosen to be parallel to the grating corrugation direction, and the  energy is set at \qty{1.61}{\milli \electronvolt}.
The excitation beam is focused into a \qty{3.90(0.03)}{\micro \meter} $1/e^2$ spot through a NA = 0.68 aspheric objective, at the centre of the etched structure.

\textbf{Interferometric measurements.}
The photoluminescence collected by the objective is sent into a Michelson interferometer through the detection line.
In one of the two arms the image is centro-symmetrically flipped along $x$ and $y$ using a backreflector and the coordinate-space photoluminescence distribution re-focused onto the spectrometer slits through the same optical path.
The total magnification of the line, sample-to-camera, is approximately 72.6.
The two arms are synchronised and balanced, optimising the visibility of the interference pattern of the excitation laser reflected from the sample surface.

\textbf{Reciprocal space measurements}
For the reciprocal space images, a similar detection line is used, but the interferometer is bypassed.
A 4f-system collects the reciprocal plane and focuses it onto the monochromator slits, resulting in a total magnification of the objective back focal plane equal to 1.66.

\textbf{Coherence calculation.}
The two-point correlator $g^{(1)}(x-x')=\braket{E^*_1(x)E_2(x)} / \braket{|E_1(x)|^2}\braket{|E_2(x)|^2}^{1/2}$, where $E_{1,\,2}(x)$ are the fields from the two interferometer arms, is proportional to the visibility $\mathcal V(x)=(I_u - I_l) / (I_u + I_l)$ of the interferometric fringes, where $I_u$ ($I_l$) is the upper (lower) envelope of the interferogram; we omit the $x$ dependence for brevity.
We calculate $\mathcal V$ by first extracting the $k=k_r$ and $k=0$ components of the interferogram by appropriate filtering in the reciprocal space and then using the Hilbert transform to calculate the envelope of the ac component; the visibility is then simply the envelope over the $k=0$ component.
For slight imbalance or not full reflection symmetry between the two images, $\mathcal V$ needs to be rescaled by $2\sqrt{I_1I_2}/(I_1 + I_2)$, where $I_{1,2}$ are the intensities in each arm of the interferometer, to give a correctly normalised $g^{(1)}(x-x')$.
We forgo this normalisation since small misalignments of the order of $1/k_r$ in the overlap of the two images distorts the density modulation.
This does not significantly alter the form of $g^{(1)}(x-x')$ since in our case this prefactor is $\simeq 1$.
\smallskip

\textbf{Acknowledgements.}
We acknowledge fruitful discussions with V. Ardizzone, M. Pieczarka, and A. Recati.
We are thankful to G. Lerario for his valuable feedback on the manuscript.
This project was funded by the Italian Ministry of University (MUR) PRIN project ``Interacting Photons in Polariton Circuits'' – INPhoPOL (grant 2017P9FJBS);
the project ``Hardware implementation of a polariton neural network for neuromorphic computing'' – Joint Bilateral Agreement CNR-RFBR (Russian Foundation for Basic Research) – Triennal Program 2021–2023;
the MAECI project ``Novel photonic platform for neuromorphic computing'', Joint Bilateral Project Italia - Polonia 2022-2023;
the PNRR MUR project: ``National Quantum Science and Technology Institute'' -  NQSTI (PE0000023) co-funded by the European Union - NextGeneration EU;
the Apulia Region, project ``Progetto Tecnopolo per la Medicina di precisione'', Tecnomed 2 (grant number: Deliberazione della Giunta Regionale n. 2117 del 21/11/2018);
the PRIN project ``QNoRM: A quantum neuromorphic recognition machine of quantum states'' - (grant 20229J8Z4P).
This research has been cofunded by the European Union - NextGeneration EU, ``Integrated infrastructure initiative in Photonic and Quantum Sciences'' - I-PHOQS [IR0000016, ID D2B8D520, CUP B53C22001750006].
IC acknowledges funding from the Provincia Autonoma di Trento, partly via the Q@TN initiative.
This research is funded in part by the Gordon and Betty Moore Foundation’s EPiQS Initiative, grant GBMF9615 to L.P., and by the National Science Foundation MRSEC grant DMR 2011750 to Princeton University.
Work at the Molecular Foundry is supported by the Office of Science, Office of Basic Energy Sciences, of the U.S. Department of Energy under Contract No. DE-AC02-05CH11231.
We thank Scott Dhuey for assistance with electron beam lithography and Paolo Cazzato for the technical support.
\smallskip

\textbf{Competing interests.}
The authors declare no competing interests.
\smallskip

\textbf{Data Availability.}
The data of this study is available from the corresponding author upon reasonable request.
\smallskip

\textbf{Author contributions.}
DT and ML conceived the experiment and convinced DS to go ahead with it.
AG took the data and together with DT performed the analysis.
DT, AG and ML wrote the manuscript.
DN, DG, and IC provided theoretical support.
FR processed the sample; growth was performed by
KB and LP.
DS supervised the work, discussed the data, and tirelessly explained to DT how polariton OPOs should behave.
DT tried to use the best photonics language he could muster but in the end had to use kets.
All authors contributed to discussions and editing of the manuscript.

\putbib
\end{bibunit}

\clearpage
\newpage

\begin{bibunit}

\begin{center}
  \textbf{\large Supplementary Information}
\end{center}

\setcounter{equation}{0}
\setcounter{figure}{0}
\setcounter{table}{0}
\setcounter{section}{0}
\setcounter{page}{1}
\renewcommand{\theequation}{S\arabic{equation}}
\renewcommand{\thefigure}{S\arabic{figure}}

\section{Dispersion of the waveguide}

The particle-number-conserving part of the system Hamiltonian can be written as a $8\times 8$ matrix
\begin{equation}
\renewcommand{\arraystretch}{1.2}
\definecolor{light-gray}{gray}{0.8}
\arrayrulecolor{light-gray}
  \mathcal H(k) = \left(\begin{array}{cc|cc|cc|cc}
    \color{blue} \delta_0 & \color{blue} U_{00} & V_0 & 0 & 0 & 0 & 0 &U_{01}\\
    \color{blue} U_{00} & \color{blue} -\delta_0 & 0 & V_0 & 0 & 0 & U_{01} & 0  \\\hline
    V_0 & 0 & 0 & 0 & 0 & 0 & 0 & 0  \\
    0 & V_0 & 0 & 0 & 0 & 0 & 0 & 0 \\\hline
    0 & 0 & 0 & 0 & 0 & 0 & V_1 & 0  \\
    0 & 0 & 0 & 0 & 0 & 0 & 0 & V_1 \\\hline
    0 & U_{01} & 0 & 0 & V_1 & 0 & \color{red} \delta_{1} &  \color{red} U_{11} \\
    U_{01} & 0 & 0 & 0 & 0 & V_1 & \color{red} U_{11} &  \color{red} \delta_{-1}
  \end{array}\right),
  \label{eq:blockH}
\end{equation}

which is represented on the basis $\{\ket{\pm 0},\ket{X}^4,\ket{\pm 1}\}$ comprising the four TE$_0$ and TE$_1$ co- and counter-propagating modes and the exciton which couples independently to left- and right-propagating modes of different momenta.
Here we take the energy of the bare exciton resonance as a reference, i.e., we set it to zero.
The linear dispersions of the modes $\delta_j\equiv\delta_j(k)=\delta\pm u (k - j k_r)$, where $j=0,\pm 1$, $\delta$, act as effective detunings that are identical up to a translation in $k$; $\delta$ is the photonic detuning, $u$ the photonic group velocity, and the (plus) minus sign is for (right) left-propagating modes.
The photonic modes are coupled with Rabi frequencies $U_{ij}$, that depend on the mode overlap with the etched grating, and then couple to the degenerate exciton mode with Rabi frequencies $V_{0,1}$, with $V_1 < V_0$ due to the different mode distributions of $\ket{0}$ and $\ket{\pm 1}$.
The block structure of the Hamiltonian makes it straightforward to identify the contributing parts.
The photonic part of the BiC can be described with a $2\times 2$ matrix and is analogous to a spin-$1/2$ system (top left block).
The counter and co-propagating TE$_1$ modes exist at the same energy as the TE$_0$ modes and define the period of the supersolid (bottom right block).

The BiC state can be simply described as a dissipative spin-$1/2$ system
\begin{equation}
  \mathcal H_B(k) = \begin{pmatrix}
    \delta_0 & U_{00}\\
    U_{00} & -\delta_0
  \end{pmatrix} + i\gamma_0 \begin{pmatrix}
    1 & -1\\
    -1 & 1
  \end{pmatrix},
\end{equation}
with eigenvalues $\lambda_{1,2}=-i\gamma_0 \pm \sqrt{\delta_0^2 + (U_{00} - i\gamma_0)^2}$.
The near-zero linewidth of the BiC stems from the fact that at $\delta_0=a k = 0$ one of the eigenvalues is purely real (and corresponds to the BiC state) while the other one has twice the dissipation $2i\gamma$.

\section{Resonant excitation of propagating modes}

\begin{figure}[htb]
	\begin{center}
	 \includegraphics[width=\columnwidth]{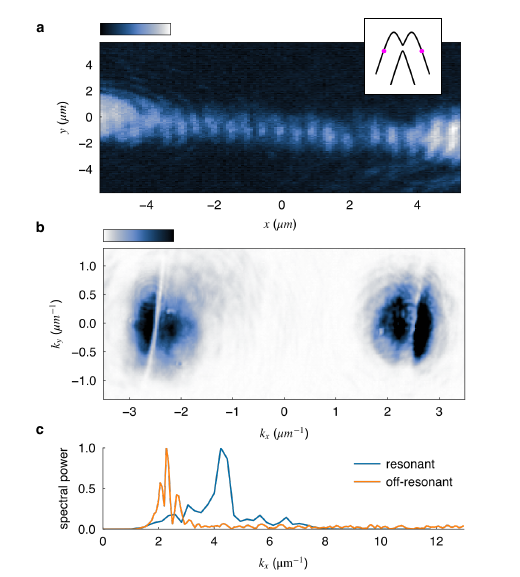}
	\caption[]{\textbf{Resonant excitation of propagating modes.} \textbf{a.} Two resonant spots injected at the energy of the BiC interfere and give rise to an optical lattice structure. \textbf{b.} The spots enter resonantly at $\ket{\pm 1}$ and propagate towards $k=0$. The interference pattern, shown in \textbf{c.} has twice the spatial frequency of that of the supersolid (off-resonant) since $\ket{\pm 0}$ is absent. The supersolid peak is split due to the non-connected spatial structure of the BiC.}
	\label{fig:s1}
	\end{center}
\end{figure}

By resonantly exciting the TE$_{\pm 1}$ modes at the energy of the BiC we effectively isolate the ``solid'' part of our polaritonic supersolid and measure its characteristics.
The interference that arises from the diffracted peaks has $k=2k_r$.
Note that this in contrast with results discussed in the main text. 
Indeed, since in the present case the total emitted field is the superposition of two plane-wave contributions counter propagating, without any signal coming from TE$_0$, the field intensity is $\cos(k_r \cdot x)^2\simeq \cos(2k_r \cdot x)$, resulting in a $k=2k_r$ peak in the emitted spectrum.
To avoid interference coming directly from the laser we use two distinct excitation spots for $\pm k_r$ at a distance of approximately \qty{10}{\mu m} and measure the interference of the propagating modes.
Figure~\ref{fig:s1}a shows the polariton density in coordinate space normalised to the density of the individual laser beams; the interference in the region between the spots is evident.
Both lasers enter at $\ket{\pm 1}$ (see \cref{fig:s1}b) and due to the negative group velocity propagate towards $k=0$.
Notice that the fringes do not have an apparent curvature which is characteristic of the phase fluctuations due to the spontaneous symmetry breaking of the supersolid.
Moreover, the Fourier transform has a single well-defined peak since the gaussian beams have a slowly-varying spatial structure.
Interference between the BiC and $\ket{\pm 1}$ on the other hand, shows multiple peaks around $k_r$ corresponding to the two BiC lobes (see \cref{fig:s1}c).

\section{Linear coupling and finite size effects}

Since we are imaging an ensemble average of many supersolid realisations it might seem surprising that we observe the breaking of translational symmetry in the first place.
Should this phase transition be truly spontaneous, the phase of the density modulation would average to a uniform distribution in $[0,2\pi]$ and significantly reduce the visibility of the fringes.
As explained in the main text, the observed residual modulation is due to a symmetry breaking term resulting from the linear coupling of the modes in our system so we would not recover full phase freedom even at the thermodynamic limit unless $U_{01}\to 0$.
We note that in the thermodynamic limit of very large pump power, the nonlinear coupling scales more favourably with $n_0$ than the linear coupling and our system converges to an ideal supersolid.
For vanishing linear coupling, the next-order, finite-size effects are the trapping potential from the laser and the negative effective mass~\cite{riminucciPolaritonCondensationGapConfined2023}, and the length of the waveguide.
The length of the waveguides in this work is \qty{300}{\mu m} which is comparable to the propagation length of the BEC and as such it reaches the edges.
However, we did not observe any differences using \qty{1}{mm}-long waveguides with tapered edges that should minimise reflections, which means that such standing waves do not appear in our system.

\subsection{Spatial phase shift}

\begin{figure}[htb]
	\begin{center}
	 \includegraphics[width=\columnwidth]{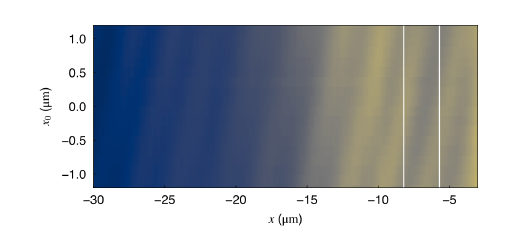}
	\caption[]{\textbf{Phase rolling.} The phase of the density modulation changes smoothly when varying the position $x_0$ of waveguide along $x$ keeping the pump laser fixed. The phase winds over $w\pi$ for a rigid translation of $k_r^{-1}$ indicating the pointing stability of the system.}
	\label{fig:s21}
	\end{center}
\end{figure}

When scattering from a moving grating the scattered orders pick up a phase factor $\propto e^{imk_G(x-x_0)}$, where $m$ is the diffraction order, $k_G$ the grating wavevector, and $x_0$ the position of the grating with respect to an arbitrary origin~\cite{wise.PhysRevLett.95.013901}.
In our case the origin is defined by the relative position of the laser spot and the grating.
Figure~\ref{fig:s21} shows the density modulation changing with the centre position of the excitation spot $x_0$ along the waveguide axis.
The density modulation of our BEC can be considered as two plane waves impinging onto the grating.
If there is relative translation of the grating or the laser spot the modulation acquires a phase shift proportional to $k_rx_0$.
In our case, these small spatial fluctuations are negligible since for a rigid translation of $x_0=k_r^{-1}=\qty{2.4}{\mu m}$, much smaller than the extent of the BEC, the phase winds over 2$\pi$.

\subsection{Transverse phase fluctuations}

\begin{figure}[htb]
	\begin{center}
	 \includegraphics[width=\columnwidth]{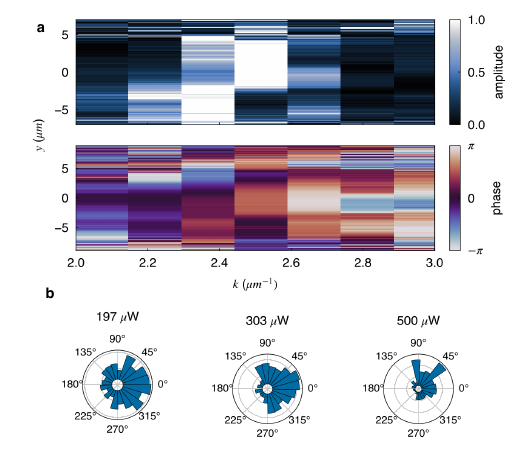}
	\caption[]{\textbf{Transverse phase variations.} \textbf{a.} Amplitude and phase of the spatial frequencies of the BEC density. The periodicity of the crystal remains largely constant when the intensity of the BiC is large enough, however the maxima of the dislocations fluctuate for $|y| > \qty{2.5}{\mu m}$. \textbf{b.} The phase distribution, at a fixed periodicity $k_r$, narrows with increasing power spanning the full unit circle below threshold to about \qty{135}{\degree} at \qty{500}{\mu W}.}
	\label{fig:s22}
	\end{center}
\end{figure}

A degree of phase freedom can be inferred by the transverse phase variations of the density modulation.
The two-dimensional density distribution is proportional to $1 + |a(y)|^2 \cos{\big(k_r x + \theta(y)\big)}$, where $a(y)$ and $\theta(y)$ are the transverse amplitude and phase of the density modulation.
Figure~\ref{fig:s22}a shows the reciprocal amplitude and phase of a typical BEC realisation.
When $|y| < \qty{2.5}{\mu m}$ the peak maxima that define the periodicity of the crystal are at $k_r$ as expected.
Fluctuations increase for larger values of $y$ as the density drops.
Compare this with \cref{fig:s1} where $\ket{\pm 1}$ are excited resonantly and the interference pattern does not show any transverse variation.

The transverse phase of the crystal structure $\theta(y)$ measured at $k_r$ is uniformly distributed in~\cref{fig:s22}b for low powers (left panel) when the system is still largely incoherent.
The distribution narrows with increasing pump power till it is mostly distributed over the first quadrant at \qty{500}{\mu W} (right panel).
Critical fluctuations at lower densities, close to threshold, likely scramble the U(1) phase which is then pinned at higher densities.

\subsection{Average phase fluctuations}

\begin{figure}[htb]
	\begin{center}
	 \includegraphics[width=\columnwidth]{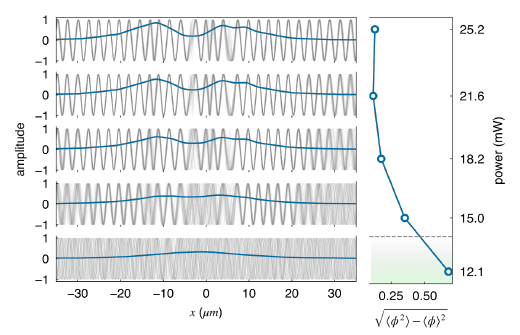}
	\caption[]{\textbf{Average phase fluctuations.} The amplitude modulation for 50 different continuous-wave realisations of the supersolid (gray). As the pump power goes above threshold the fluctuations are mainly visible in the minima and maxima of the density. The standard deviation of the phase (right panel), measured only at the maxima, reduces above threshold and seems to increase at a power of about $\times 1.5$ the threshold power.}
	\label{fig:s23}
	\end{center}
\end{figure}

Here, we switched to continuous pumping in order to avoid averaging from the pulsed laser as described in the main text.
Figure~\ref{fig:s23} shows 50 BEC realisations for powers around the condensation threshold which in this case is around \qty{14}{mW}.
Below this threshold the phase of the modulation is completely random since the system is still incoherent, and its standard deviation is about 0.7.
The standard deviation is measured only in the region of \qty{5}{\mu m} to \qty{10}{\mu m} to avoid artefacts of fluctuations coming from near-zero density.
The standard deviation reduces monotonically till about $\times 1.5$ the threshold power and saturates or slightly decreases again, indicating that the phase is yet to be fully fixed in the high density regions.
This measurement is not sensitive to pointing fluctuations of the laser beam since it shows local phase fluctuations only.

\section{Holographic imaging}

\begin{figure}[htb]
	\begin{center}
	 \includegraphics[width=\columnwidth]{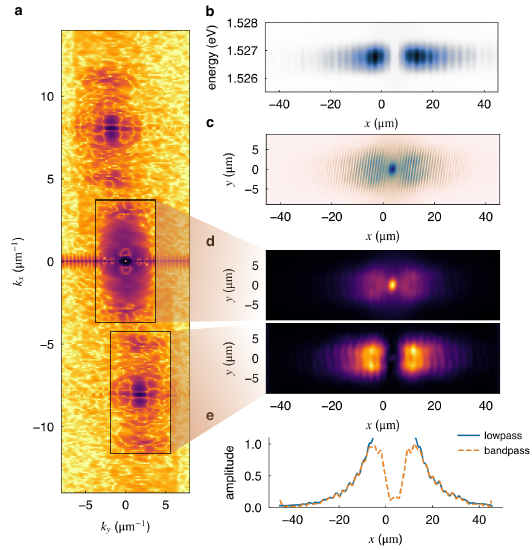}
	\caption[]{\textbf{Holographic imaging}. \textbf{a.} Fourier transform of the interferogram in \textbf{c.} showing the main peak around \qty{7.5}{\mu m^{-1}} and sidebands at $\pm k_r$ from it. \textbf{b.} In energy resolved measurements the BiC is dark in the middle.\textbf{c.} An interferogram as given by the Michelson interferometer described in Methods. The bright spot in the middle is coming directly from the exciton reservoir and its size is comparable to the laser spot. \textbf{d.} Retaining only the middle frequencies leads to a density without the middle spot (bottom panel) since it is incoherent, as opposed to retaining the low frequencies (top panel). \textbf{e.} In both cases, the amplitude of the modulation and the density in the lobes of the BiC is exactly the same.}
	\label{fig:s4}
	\end{center}
\end{figure}

The small excitation spot creates a hot exciton reservoir underneath it that also emits photolumiscence covering the BiC notch at $k=0$ (see \cref{fig:s4}c and d).
This feature does not exist in energy-resolved measurements, as shown in \cref{fig:s4}a.
By taking advantage of interferometric techniques we can remove it also in energy-averaged measurements and recreate a realistic representation of the density in coordinate space.
The reciprocal-space representation of the interferogram (\cref{fig:s4}a) shows, apart from the low-lying wavevectors, a strong peak at $k_2 - k_1=\qty{7.5}{\mu m^{-1}}$ which corresponds to the fringe spacing from the interferometer.
The visible sidebands on the fringe spacing are at $\pm k_r$ and arise from the beating of the density modulation with the interferometer fringes.
The fringe spacing has been chosen to be $\approx 3k_r$ so that the lowest sideband at $k_2 - k_1 -k_r$ is above all the low-lying wavevectors.
Using a bandpass filter before transforming back to coordinate space allows us to selectively eliminate the incoherent central spot while otherwise leaving the density unaffected (see \cref{fig:s4}d and e).

\putbib
\end{bibunit}

\end{document}